\documentclass[prl,preprint,amsmath,amssymb,superscriptaddress,showpacs]{revtex4}

\usepackage{graphicx}
\usepackage{dcolumn}
\usepackage{bm}
\usepackage[sort&compress]{natbib}

\begin{document}


\title{Externally Mode-matched Cavity Quantum Electrodynamics with Charge-tunable Quantum Dots}

\author{M.~T. Rakher} \email{mrakher@nist.gov}
\affiliation{Department of Physics, University of California Santa
Barbara, Santa Barbara, California 93106, USA}
\altaffiliation{present address: National Institute for Standards
and Technology, Gaithersburg, Maryland 20899, USA}
\author{N.~G. Stoltz}
\affiliation{Materials Department, University of California Santa Barbara, Santa Barbara, California 93106, USA}
\author{L.~A. Coldren}
\affiliation{Materials Department, University of California Santa Barbara, Santa Barbara, California 93106, USA}
\affiliation{ECE Department, University of California Santa Barbara, Santa Barbara, California 93106, USA}
\author{P.~M. Petroff}
\affiliation{Materials Department, University of California Santa Barbara, Santa Barbara, California 93106, USA}
\affiliation{ECE Department, University of California Santa Barbara, Santa Barbara, California 93106, USA}
\author{D. Bouwmeester}
\affiliation{Department of Physics, University of California Santa Barbara, Santa Barbara, California 93106, USA}
\affiliation{Huygens Laboratory, Leiden University, P.O. Box 9504, 2300 RA Leiden, the Netherlands}

\date{\today}

\begin{abstract}
We present coherent reflection spectroscopy on a charge and DC Stark
tunable quantum dot embedded in a high-quality and externally
mode-matched microcavity.  The addition of an exciton to a
single-electron charged quantum dot forms a trion that interacts
with the microcavity just below strong coupling regime of cavity
quantum electrodynamics.  Such an integrated, monolithic system is a
crucial step towards the implementation of scalable hybrid quantum
information schemes that are based on an efficient interaction
between a single photon and a confined electron spin.

\end{abstract}


\maketitle

Hybrid quantum information schemes combine the coherence properties
and ease of manipulation of photons with the scalability and
robustness of local quantum systems.  Examples of local quantum
systems include electron spins in quantum dots, defect centers in
diamond, and trapped atoms or ions \cite{HarocheRev,MabuchiSci02}.
Hybrid schemes such as quantum repeaters and quantum networks use
the coupling between the local quantum system (qubit) and the
optical field to reversibly map the quantum state of an injected
photon onto the state of the local system
\cite{VanEnkPRL97,CiracPRL97,YaoPRL04,YaoPRL05}.  Other hybrid
schemes use a joint measurement of emitted photons, which are
entangled with their respective local qubit, to perform gate
operations on the two spatially separated local systems
\cite{RaussendorfPRL01,LimPRL05,BarrettPRA05,SimonPRB07}.  This
latter scheme can be used to create entanglement between many local
qubits as needed for cluster state quantum computation.
 Implementations based on trapped ions or atoms have reached operation
fidelities greater than 80\% for 2 remote qubit interactions.
However, the overall success probabilities are currently limited to
$\approx$10$^{-8}$, due to the technical incompatibility of trapping
the particles and coupling them efficiently to a single external
optical mode \cite{MoehringNat07,RosenfeldPRL07}.  Here we present a
solid-state system that integrates a trapped,
electrically-controlled quantum system with near unity coupling
efficiency to an external optical mode.

To achieve an efficient coupling, the quantum system must be placed
in a high-quality microcavity so that it dominantly interacts with a
single optical mode.  Furthermore, this cavity mode must be
mode-matched to an external mode to ensure efficient operation at
the single photon level.  The ideal operating point for such hybrid
schemes is deep in the weak coupling (Purcell) regime of cavity
quantum electrodynamics (QED), just below the onset of strong
coupling. In addition, for cluster-state and distributed quantum
computation, the hybrid system must be scalable.  Our system
satisfies these requirements in the solid-state.  It is composed of
self-assembled quantum dots (QDs) (density $\approx 10 \mu m^{-2}$)
at the center of an oxidation-apertured micropillar cavity with
integrated doped layers that enable an external bias to apply an
electric field across the QD.   This field causes carriers to tunnel
in and out \cite{WarburtonNat00} of the QD, changing the QD charge
state, and induces the quantum confined Stark effect
\cite{FryPRL00,KrennerPRL06}, shifting the emitted photon's energy.
While cavity QED has been studied using quantum dots for several
years, this has been done using a neutral exciton
\cite{KartikNat07,EnglundNat07}.  Neutral excitons, bound
electron-hole pairs, have been proposed as qubits \cite{ChenPRL01}
but seem problematic due to their quick spontaneous decay, $\approx$
1ns in GaAs, and fast dephasing \cite{KammererAPL02}.  The local
qubit in our system is the spin of a trapped electron
\cite{KhaetskiiPRL02,MerkulovPRB02}, which interacts with the cavity
mode through the addition of an exciton, forming a short-lived trion
state.  Since the polarization of the emitted photon is correlated
with the spin state of the remaining electron, the trion acts as a
readout channel of the spin \cite{YaoPRL04}.  Additionally, the
micropillar cavity geometry is such that the fundamental mode is a
doubly degenerate HE$_{1,1}$ mode, which mode-matches well to
external modes due to its Gaussian-like shape \cite{PeltonIEEE02}.
Here, we report on two variations of the solid-state cavity QED
system, one optimized to operate in the charge-tuning regime, and
the other in the Stark-tuning regime.

\begin{figure}
\centering
        \includegraphics[width=8.5cm, clip=true]{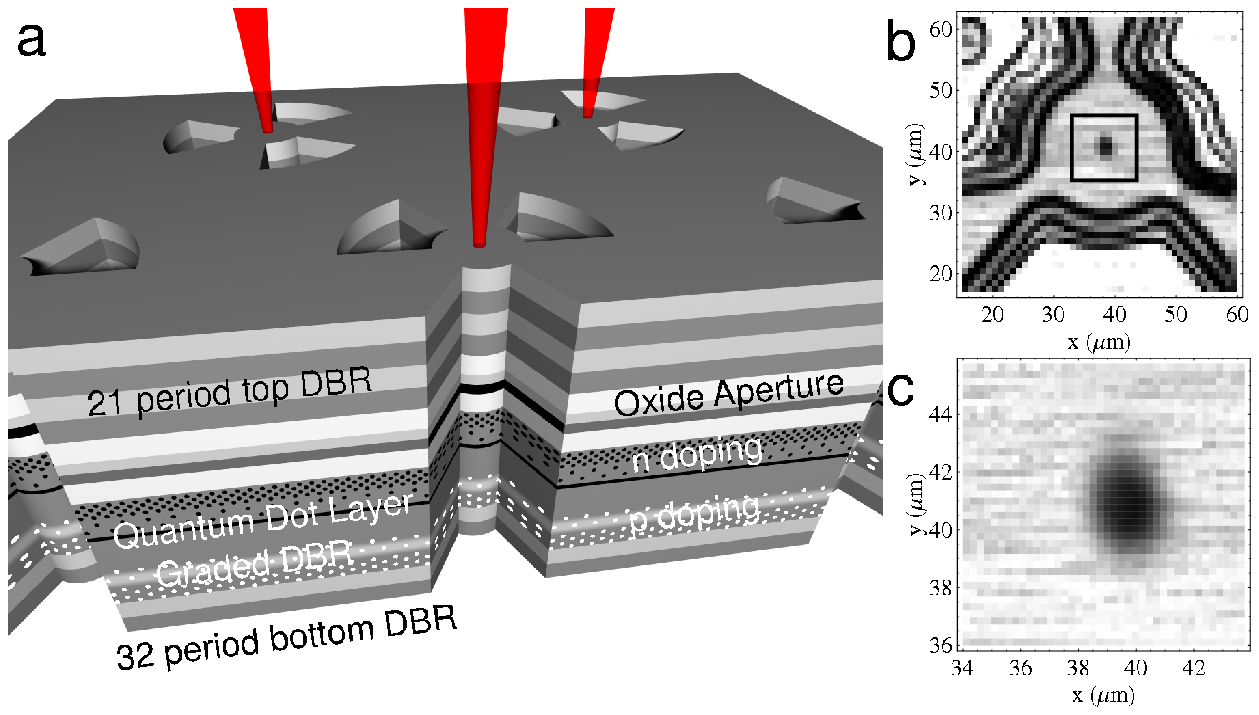}
    \caption{(a) Schematic of the scalable solid-state cavity QED system based on electrically gated self-assembled QDs embedded in oxide-apertured micropillars.(b) Two-dimensional reflectivity scan of a micropillar cavity taken with a laser resonant with the cavity mode.  The mode in the center can be seen clearly as a dip in the reflected signal. (c) Higher resolution reflectivity scan taken in a 10 $\mu$m by 10 $\mu$m area containing the mode, as depicted in (b).}
    \label{fig:fig1}
\end{figure}

The demonstration of an electrically-gated QD embedded at the
anti-node of a high Q cavity mode has become feasible through a
series of scientific advances.  Firstly, the development of
vertical-cavity surface-emitting lasers (VCSELs) with oxide
apertures in the GaAs/AlGaAs material system enabled the creation of
cavities with small mode volumes, $V_{eff} = 35 (\lambda/n)^3$,
while maintaining a very high $Q$ \cite{Coldren}.    Secondly, the
addition of single, self-assembled InAs/GaAs QDs embedded at the
axial anti-node of the cavity mode provided an atomic-like emitter
to couple to the optical mode \cite{StoltzAPL05}.  Thirdly, the use
of etched trenches to define the oxidation front, as shown in
Fig.~\ref{fig:fig1}, enabled both control over the polarization
degeneracy of the cavity modes as well as global electrical
connection to an array of solid-state cavity QED systems
\cite{StraufNphot07}.   In the experiments presented here, the
oxidation time of the aperture is such to maximize the
quality-factor, $Q$, while minimizing the mode volume.  As shown in
Fig.~\ref{fig:fig1}b,c, the cavity mode is to a good approximation
Gaussian in lateral profile and fits to a waist of 2.2 $\mu$m, in
agreement with measurements of the spacing between different lateral
modes \cite{StoltzAPL05}.

Using a voltage source to create an electric field which properly
drops over the QD active region is complicated by the presence of
nearby material interfaces at each distributed Bragg reflector (DBR)
period and at the oxide aperture region.  These interfaces trap
charges and result in the formation of charge domains, which reduce
the field dropped across the QD region and hence obstruct controlled
charging and Stark tuning \cite{StraufNphot07}.  To overcome these
problems, a novel P-I-N device structure was developed in which the
intrinsic region does not include the oxide aperture and the nearby
p-doped Al$_{0.9}$Ga$_{0.1}$As DBR period is Al-content graded to
and from the adjacent GaAs layers as shown in Fig.~\ref{fig:fig1}a.
The Al-content grading prevents the formation of triangular
potential wells that arise at abrupt Al$_{0.9}$Ga$_{0.1}$As/GaAs
interfaces.  Furthermore, all doping concentrations are graded such
that the doped regions are easily contacted by countersink etching
without introducing unnecessary dopants near the QD region.  The two
variations of the solid-state cavity QED system presented here have
nominally the same growth structure, but the average doping levels
for the charge tuning system are $3.5$ $10^{18}$ cm$^{-3}$ ($2.5$
$10^{18}$ cm$^{-3}$) for the n-doped (p-doped) layer whereas for the
Stark tuning system they are $7.0$ $10^{17}$ cm$^{-3}$ ($7.5$
$10^{17}$ cm$^{-3}$).

\begin{figure}
\centering
        \includegraphics[width=8.5cm, clip=true]{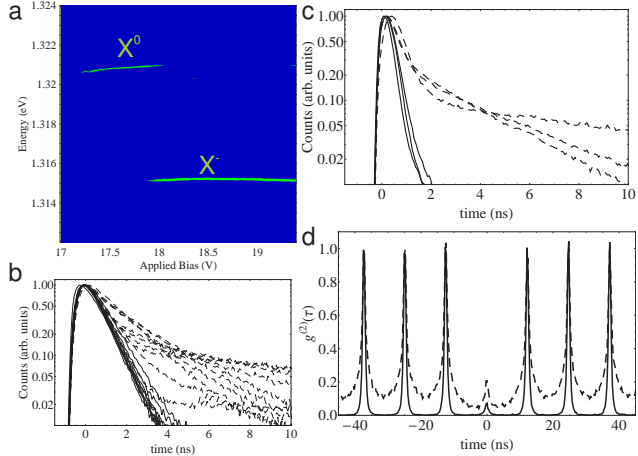}
    \caption{(a) Photoluminescence spectra as a function of applied bias for a QD in the mirror region at 4 K. (b) Emission decay traces for 10 QDs in the mirror region.  Straight traces are taken with 18.5 V applied bias ($X^-$) and dashed traces are taken with 17.5 V applied bias ($X^0$). (c) Emission decay traces for QDs on resonance with cavity modes for several different cavities.  The straight (dashed) traces correspond to an $X^-$ ($X^0$) decay.  (d) $g^{(2)}(\tau)$ measurements for an $X^-$ (straight) and $X^0$ (dashed) transition. }
    \label{fig:fig2}
\end{figure}

To investigate the intra-cavity charging, we first characterized QDs
outside of the cavities in the surrounding mirror region, where the
Purcell effect is negligible.  We monitored the photoluminescence
spectrum (using a 1.25 m monochromator coupled with a CCD array)
under 150 fs, 860 nm Ti:Sa laser excitation with 50 nW average power
as the applied bias is varied. A typical trace for a single QD is
shown in Fig.~\ref{fig:fig2}a.  Near 18 V applied bias, there is a
transition for the QD emission line to a line that is 6 meV to lower
energy.  This is the characteristic energy separation for the
transition between the neutral exciton, $X^0$, and the singly
electron-charged exciton, $X^-$ \cite{WarburtonNat00,AtatureSci06}.
To verify the charge designations as $X^0$ and $X^-$, we also
measured the time-resolved decay of the photoluminescence using an
avalanche photodiode with a time-to-amplitude converter. The result
is shown in Fig.~\ref{fig:fig2}b, with curves taken at biases below
(above) 18 V labelled as dashed (straight).  Because of the presence
of optically dark states, the $X^0$ decay traces have a distinctive
bi-exponential behavior, whereas the $X^-$ decay are single
exponentials \cite{StraufNphot07,SmithPRL05}.

The same lifetime measurement is performed for QDs in the cavity
region which are on resonance with a polarization-degenerate
fundamental mode as shown in Fig.~\ref{fig:fig2}c (this was possible
for approximately 10\% of cavities), and the results qualitatively
replicate that of the bulk QDs.  However, the effect of the high $Q$
cavity strongly reduces the emission lifetime by the Purcell effect.
For some $X^-$ cases, this lifetime approaches the timing resolution
of our experiment, which is $\sim$150 ps and is due to the APD
timing jitter.  Nonetheless, a deconvolved lifetime of 137$\pm$21 ps
was obtained for the fastest $X^-$ transition and 321$\pm$15 ps for
the $X^0$.  This yields a Purcell enhancement, $F_p =
\tau_{o}/\tau_{cav}$, of approximately 7 for the $X^-$.  We measure
that on average ($F_p = 2.8 \pm 0.22$ for 4 $X^0$ transitions and
$5.9 \pm 0.96$ for 6 $X^-$) the Purcell enhancement is stronger for
$X^-$ than for $X^0$.  Because both transitions have similar
lifetimes in the DBR region, resulting from similar oscillator
strengths, one would expect both to have similar Purcell effects.
However, this is not found experimentally and may be due to a better
matching of the transition dipole moment to the cavity mode
polarization for the $X^-$.  This could be explained by theoretical
calculations beyond the standard techniques for calculating the
optical transitions of a QD \cite{NarvaezPRB05}.  In addition, the
second-order photon correlation function, $g^{(2)}(\tau)$, was
measured as shown in Fig.~\ref{fig:fig2}d for an $X^0$ and an $X^-$
transition.  While both clearly demonstrate single photon behavior,
the $X^-$ is much cleaner due to its fast, single exponential decay.
The measured single photon ($g^{(2)}(0) < 0.25$) count rate was
typically $3$ $10^6$ $s^{-1}$ for an $X^-$ with an 80 MHz pump rate,
yielding a 25$\%$ extraction efficiency for the QD when corrected
for the optical and detection losses of the setup, which correspond
to a net efficiency of 15$\%$.

\begin{figure}
\centering
        \includegraphics[width=8.5cm, clip=true]{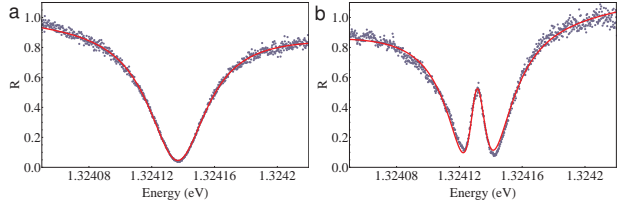}
    \caption{(a) Cavity reflection spectrum of an unloaded micropillar cavity measured by recording the reflected signal of a tunable-wavelength laser.  Eqn.~\ref{eqn:refl} with $g = 0$ plus a linear background is used to fit the data.  (b) Cavity reflection spectrum of a QD coupled to the micropillar cavity in (a).  Eqn.~\ref{eqn:refl} plus a linear background is used to fit the data.}
    \label{fig:fig3}
\end{figure}

While the lifetime measurements indicate that the emission is
coupled to the cavity mode, it does not yield a quantitative measure
of the coupling strength, $g$, or the mode-matching efficiency.  To
do this, one must probe the coupled system coherently and we
accomplish this by measuring the reflectivity of the cavity-QD
system \cite{KartikNat07,EnglundNat07}.  The reflection spectrum can
be derived from the Jaynes-Cummings Hamiltonian using the
input-output formalism and under sufficiently weak probing of a
symmetric cavity
\cite{ShenOptLett05,WaksPRL06,WaksPRA06,GarnierPRA07}, can be
expressed as
\begin{equation}
R(\omega)=\left| 1-\frac{\kappa[\gamma-i(\omega-\omega_{QD})]}{[\gamma-i(\omega-\omega_{QD})][\kappa-i(\omega-\omega_{c})]+g^2} \right|^2,
\label{eqn:refl}
\end{equation}
where $g$ is the emitter-cavity coupling, $\omega_{QD}$ ($\omega_c$)
is the emitter (cavity) resonance, $\gamma$ is the dipole decay
rate, and $\kappa$ is cavity field decay rate.  If there are no QDs
coupled to the cavity mode, the spectrum shows a single dip at the
cavity resonance with a width equal to the cavity field decay rate,
$\kappa$, as shown in Fig.~\ref{fig:fig3}a. For this micropillar, we
can fit the data to obtain $\kappa= 24.1$ $\mu$eV, which corresponds
to $Q=$27,000.  The depth of this dip is a measure of how well the
probe beam is mode-matched to the cavity, and in this case the
coupling efficiency is greater than 96$\%$.  This remarkably high
efficiency implies that reliable information transfer at the single
photon level is feasible and would constitute an increase in the
success probability of a two-photon experiment
\cite{MoehringNat07,RosenfeldPRL07} by 3-4 orders of magnitude. If a
QD is coupled to the microcavity, the reflection spectrum is
drastically altered. Figure~\ref{fig:fig3}b shows the absolute
reflection spectrum of the cavity mode interacting with a single QD
transition. By fitting this spectrum to Eqn.~\ref{eqn:refl}, we
obtain an emitter-cavity coupling of $g$ = 9.7 $\mu$eV and an
emitter decay rate of $\gamma$ = 1.9 $\mu$eV. Since $g/\kappa=0.402
$, the emitter-cavity system is deep in the Purcell (weak-coupling)
regime and at the precipice of the strong-coupling regime, $g/\kappa
> 0.5 $, exactly in the region ideally suited for hybrid quantum
information schemes \cite{YaoPRL04,YaoPRL05,LimPRL05,BarrettPRA05}.
The spectrum, with resolution limited by the probe laser linewidth,
completely characterizes the system.  Additionally, it reveals the
natural linewidth of the QD transition with a signal much greater
than achieved in transmission or differential transmission.  In
conclusion, the combination of these results for the cavity QED
system in the charge-tuning regime demonstrates that it is ideal for
hybrid quantum information processing.

\begin{figure}
\centering
        \includegraphics[width=8.5cm, clip=true]{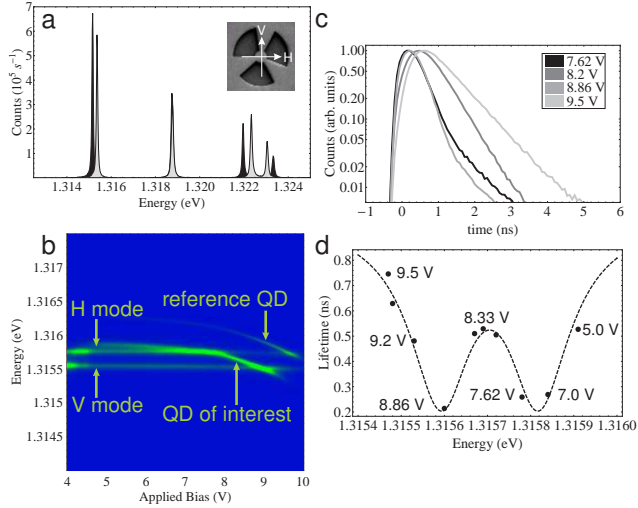}
    \caption{(a) Micro-photoluminescence spectra of non-degenerate optical modes in a micropillar.  H (V) polarized modes are black (grey).  Inset: SEM image of a micropillar.  (b) Photoluminescence spectra as a function of applied bias for two QDs (labeled QD of interest and Reference QD) and two nondegenerate fundamental cavity modes (labeled H mode and V mode).  (c) Lifetime traces for a few bias settings; 7.62 V, 8.2 V, 8.86 V, 9.5 V.  (d) Deconvolved lifetimes as a function of emission energy with fit.}
    \label{fig:fig4}
\end{figure}

We now turn to the Stark-tuning cavity QED system.  Since the
coupling between the QD and the cavity mode depends on the spectral
detuning, an external control is necessary to reach resonance.  In
QD systems without electrical gating, this control is achieved by
adjusting the sample temperature \cite{EnglundNat07}. However, this
control typically decreases coherence within the system through
higher phonon occupations and is not scalable.  An applied electric
field can also tune the QD transition via the Stark Effect without
the negative effects of temperature and in principle can be scalable
by gating each cavity separately. In order to illustrate this effect
and potential applications, we utilized a polarization
non-degenerate cavity mode. As mentioned in
Ref.~\cite{StraufNphot07}, an engineered ellipticity of the aperture
lifts the polarization degeneracy, creating two orthogonal linear
polarization modes (denoted as H and V) as illustrated in
Figure~\ref{fig:fig4}a. Because the $Q$ factor is very high
(40,000), the modes can be spectrally separated by as little as 50
$\mu$eV and still be resolved.  This enables the quantum dot
transition to be Stark-shift tuned into resonance with two modes as
shown in Fig.~\ref{fig:fig4}b.  Note that the dependence is
nonlinear with bias as expected for the Quantum Confined Stark
Effect \cite{FryPRL00,KrennerPRL06}.

By Stark-shift tuning the QD emission, a polarization-dependant
Purcell effect is observed on resonance with each mode.  Stark shift
tuning as opposed to current induced heating was confirmed by
observing a constant QD linewidth over the tuning range.  For
several applied biases, the QD emission decay curve is measured, see
Fig.~\ref{fig:fig4}c, and the extracted lifetime is plotted as a
function of spectral position as shown in Fig.~\ref{fig:fig4}d.  The
dips in the transition lifetimes measured on resonance are a clear
consequence of the Purcell effect.  The lifetimes at the resonance
of each mode is measured to be around 220 ps as shown in
Fig.~\ref{fig:fig4}d, approximately 5 times shorter than the bulk
lifetime.  The appearance of a bi-exponential, most prevalent for
the on resonance biases, in Fig.~\ref{fig:fig4}c is attributed to a
small fraction ($\sim$4\%) of photons collected from QDs outside the
mode volume.  Stark-shift tuning when used in addition to charge
tuning constitutes a completely bias-controlled, solid-state cavity
QED system.

In conclusion, we presented a solid-state cavity QED system which
has near ideal properties for photon electron-spin coupling as
needed for hybrid quantum information processing.  The unique
features of our system are intra-cavity electron charging, near
perfect mode-matching, polarization control of the cavity modes, and
operation deep in the Purcell regime.  In addition, the cavity-QD
coupling can be controlled via the Stark Effect, which has
applications for quantum and classical communication.  The
combination of this work with spin initialization, manipulation, and
readout \cite{AtatureSci06,GerardotNat07,DuttPRL05,MikkelsenNPhys07}
as well as techniques for active positioning of quantum dots
\cite{HennessyNat07} will bring the implementation of solid-state
hybrid quantum information protocols within reach.

The authors would like to thank M.~P. Van Exter and W. Yao for useful discussions.  This work was supported by NSF NIRT No. 0304678, DARPA No. MDA 972-01-1-0027, and Marie-Curie EXT-CT-2006-042580.  

\end{document}